# Computable Domains of a Halting Function

by Abel-Luis Peralta

Abstract:

We discuss the possibility of constructing a function that validates the definition or not definition of the partial recursive functions of one variable.

This is a topic in computability theory, which was first approached by Alan M. Turing in 1936 in his foundational work “On Computable Numbers”. Here we face it using the Model of computability of the recursive functions instead of the Turing’s machines, but the results are transferable from one to another paradigm with ease. Recursive functions that are not defined at a given point, correspond to the Turing machines that “do not end” for a given input. What we propose Is a slight slip from the orthodox point of view: the issue of the self-reference and of the self-validation is not an impediment in imperative languages.

## § 1. Definable versus calculable: principle of uncertainty.

One of the problems that haunted mathematicians in the first decades of the twentieth century was to recognize when a well-defined function was actually calculable, because being properly defined does not ensure that it is effectively calculable. A widely used example is that of any sentence in the formal language of Peano’s arithmetic. Given a deduction, it is easy to verify if it is correct or not, and therefore, to know if that sentence is a theorem or not, following the rules of inference of the first order logic. Thus, we could say that the set of theorems is formally well defined. But there is no procedure that allows us to decide whether a sentence (or its negation) has a demonstration. We have procedures to verify if something is a demonstration, but not to find a demonstration from scratch.

The result of these efforts were the recursive functions defined by Kurt Gödel [1931], within the framework of his incompleteness theorems; Church’s thesis, and the theorem of the non resolvability of the Halting Problem (Turing [1936]).

Since the most notorious results were negative, it remains to investigate how much we can do positively, within the limits that these results have marked us. What we will try to convey is that, despite the spectacularity of the limiting events, what they leave out of our reach are very specific and restricted issues, linked to epistemological and meaning issues rather than difficulty or impossibility of calculation. We propose to start considering the Halting Problem from the point of view of the computational model of recursive functions, because although Kleene demonstrated that the three computational models (recursive functions, Church’s lambda calculus, and Turing machines, in historical order) are comparable in many respects, recursive functions allow us to approach the extensional aspect of functions (such as ordered n-tuples) rather than the intensional aspect (such as lists of rules of calculation). It is the hard core of functions, and in a sense, if we are allowed to compare, like material substances in the physical sciences. It is so also because they have a spatial model, without resorting to geometry, if we imagine them as three-dimensional arrays of matrices.

It may happen that we have a very precise rule for deciding whether a number is a member of a set or not; but that we don’t know if there is any number that complies with the rule (Fermat’s last theorem, until Andrew Wiles’s 1993/95 demonstration, Goldbach’s conjecture to this day). This usually happens when we start from an easily calculable function (for example a polynomial of a single variable), but then we define another function, based on the first, in inverse way. Since such an equation may have no solutions, the function is not directly calculable at all points.

$f(x) = a_1 . x + a_2 . x^2 + a_3 . x^3 + \ldots + a_n . x^n$

$g(y) = \mu x [f(x) = 0 \ \& \ x > y]$ (μx is the minimum value of x that satisfies the equation)

In this example, f(x) is directly calculable for any x. While g(y) may not be defined in some values of y, because it depends on an inverse calculation of f, and on the existence of integer roots in the equation. And it may even happen that we cannot know by direct calculation whether it is not defined.

The problem of the inverse values of a function is what algebra solved for centuries, but in the past the subject was not considered in these terms: as a function that is defined but not actually calculable.

We will define the recursive functions as follows.

A function is primitive recursive if defined from the basic functions: projection (defined on an n-tuple and an index, returns the component of the n-tuple corresponding to the index), constant (always returns the same natural number for any argument), and successor; or by combining other recursive functions (others already defined according to the previous rules), applying the rule of composition (Gödel [1931]):

$h(x) = f(g(w))$

… Or the recursion rule:

$\varphi_i(0) = p$
$\varphi_i(x+1) = g(q, \varphi_i(x))$

It can also be defined indirectly (Kleene[1936]):

$\varphi_i(x) = \mu z[\, f(x, z) = 0 \,]$

... where f (x, z) is recursive primitive, and μ symbolizes the minimum value of z satisfying the equation, if any. If there is not, it is simply not defined for the value of x considered. In this case it is said to be a partial recursive function. If a function is defined indirectly, may be we do not know if it is defined for all natural numbers, we will say that it is general recursive if we can narrow the domain. Otherwise, it would not be recursive.

Suppose we give an order to the recursive functions. Only if the functions $\varphi_i(x)$ were all primitive recursive, there is a procedure to decide if the function is defined in its argument, but if they are general recursive, we do not always have a procedure to know if there is any solution of the internal equation. Then, we have the following possible situations (let's call them degrees of uncertainty):

1. $\varphi_i(x)$ is primitive recursive, then $\varphi_i$ is defined for all x, since it is total.

2. $\varphi_i(x)$ is general recursive, and we can show that there is a solution for the equation at $x=x_0$, then $\varphi_i(x_0)$ is defined at that value.

3. $\varphi_i(x)$ is general recursive, and we can show that there is no solution for the equation at $x=x_0$, then $\varphi_i(x_0)$ is not defined at that value (partial recursive).

4. $\varphi_i(x)$ is general recursive, and we cannot prove that there is a solution to the equation, nor that there is no solution at the point $x_0$. Then we don't know what value has $\varphi_i(x_0)$, because by traversing all the numerical succession for z, at some point, we could find a solution for ***f(x₀, z) = 0*** (of that elementary method is treated when we have a μ operator) and in that case would be defined; or it could be that the verification never ends, and in that case it would be undefined, but we could never guarantee it.

In the last case, we can't know if there is any pair in the set of ordered pairs <x, y> in the i th function place (matrix number i) that x = $\boldsymbol{x_0}$, in the current situation of mathematical knowledge in our culture. If we assume mathematical realism (a classical, non-constructive viewpoint), $\varphi_i(x)$ should have a given status, because considering only the extension (the ordered pairs), it follows that either the function $\varphi_i$ is defined in a given x (there is an ordered pair whose first element is $\boldsymbol{x_0}$) or is not, independently of our current knowledge of that fact, and of the existence or otherwise of a procedure for calculating the value if any (Mendelson [1990]).

It is very controversial to consider as "definition" a rule whose development can have infinite steps; consequently, the function may have a problem of definition and not of computability. However, because the internal function (to which the minimization operator, μ, is applied) is restricted to primitive recursive functions, it is possible that the methods of numerical analysis solve this case completely, and its application does not violate the terms in which it is raised, since the analytic theory of numbers has been applied for many years. To separate the roots it suffices that we see where the function changes its sign, without doing the same the first derivative, and without having zeros the second derivative. This knowledge, which comes from differential analysis, can be translated into methods that can be expressed in integer numbers. And when the interval has only length 1, we can just try the bounds to see if the root is integer or not. So the fourth uncertainty situation is reducible to the second or third.

It is possible to affirm with totally constructive criteria that there is no limitation of the functions of second order that is not already established in the first order ones. The degrees of uncertainty we have detailed may serve as the basis for a second order procedure or function (those whose arguments are not simply natural numbers but a list of first order recursive functions or procedures), and cannot add further uncertainty since they are only a partition on an ordering of first-order functions. There is no loss in the second order: if we know something in the first order, we also know it in the second (about the functions, not in logic).

It could be thought that there is no gain either. But it's not like that. Although in the "realist" method (all ordered pairs) there would not be any gain (since exhaustive definitions may be lacking in practice), there may be gain in the syntactic analysis: we can know that certain equations have solutions and some do not, without necessity to scrutinize its ideal matrix of values.

Therefore, if TMs form a computational model essentially equivalent to recursive functions, and first-order TMs (which do not process other TMs) compute the values of first-order recursive functions, there can be no problems with order two neither for the TMs.

However, results such as those of the non-resolvability of the Halting Problem seem to confront this conclusion. We will try to answer that question in the next paragraph.

## § 2. Self-contradictory sentences versus *reductio ad absurdum*

Historically the problem of finiteness of calculations and automation of the revision of undefined values in the recursive functions is raised after Church's thesis, and its version by Turing in 1936. In fact the problem was "imported" from the Turing machines to the recursive functions by Kleene and Davis. From then, it is classic that the literature on computability includes some kind of demonstration about the impossibility of constructing a second order function that detects the states of undefinition of the first order functions.

As a previous step, let us see that given an enumeration of general recursive functions, it is always possible to construct from it, using only purely recursive operations, a function that is not recursive. The method is analogous to Cantor's diagonalization: let $<g_i>$ be the list of functions; we can construct a function ***h*** as follows:

$$h(i) = g_i(i) + 1$$

If ***h*** were recursive, then it would be in the list: there is an index i = k such that

$$g_k(x) = h(x)$$

But by the definition of h:

$$h(k) = g_k(k) = g_k(k) + 1 \Rightarrow 0 = 1 \text{ :: absurd.}$$

Therefore, h(x) cannot be in the list, and as it is assumed exhaustive, the function that is not in it is not recursive.

The non-recursive step in the construction of h is its transfinite condition: to construct h we need, besides an infinite number of functions, to locate a function that at the same time belongs to the list and transcends it, because it can not have a fixed order in the list. So, although it is used as an argument against the recursion of the construction of h, it is against the exhaustiveness of the list. In fact, for any finite subset of functions it is possible to construct such a function.

This diagonalization over general recursive functions is sometimes interpreted as an absolute impossibility to construct second-order functions to determine the ranges of first-order recursive functions. Let us see below the construction of several types of contradiction in the terms that in texts often go through "demonstrations by the absurd" to substantiate this impossibility.

2.1. Self-reference and self-application

We will call self-reference to having a sentence of a descriptive language a subject that is part of the extension of the whole sentence that contains it (i.e., the subject of p satisfies p). We will speak of self-application when in an imperative language a free variable is instantiated by the code of the open formula that contains it.

It might seem that the previous four degrees of uncertainty exhaust all possibilities. However, let us consider a function with two arguments and of second-order (whose first argument is a list of first-order general recursive functions with one argument; see Kleene[1952], § 58; Mendelson[1979], Chapter 5, Sec. 4):

$$\theta: <N \times N> \rightarrow \{0, 1\}$$

$$\theta(i, x) = \begin{cases} 1 & \text{if} \quad \exists k\ \varphi_i(x)=k \quad (\varphi_i \text{ is } \textit{defined} \text{ for } x) \\ 0 & \text{if} \quad \neg\exists k\ \varphi_i(x)=k \quad (\varphi_i \text{ is } \textit{undefined} \text{ for } x) \end{cases}$$

Let us consider a special function[1] (lets call this the diagonal formula):

$$\eth(x) = \theta(x, x)$$

Is there an index, say ***n***, such that $\varphi_n(x) = \eth(x)$? If there is, then let as consider $\eth(n)$.

Suppose the result is 1. We see that it depends on the value of another instance of itself (n is its own index). But since there is no argument that we can apply to the second instance, it is the index of an open formula. Then it should be independent of the variable, a constant function equal to 1 for any value of the argument. (In case of supposing that implicitly has as its argument its own index would generate an infinite regress, which would make the final value undefined, and the result value must be 0: contradiction). But this would mean that in cases where the argument function is undefined, the value is 1, contradicting the definition, which assigns it 0 in that case. Absurd.

Suppose the result is 0. In this case, according to the definition of θ, the function whose index is n should be undefined in n. But n is the index of ð (n); so it is undefined in n but also 0: nonsense.

Therefore, the only consistent possibility is that ð is not defined at its own index. But then neither is θ(n, n).

The usual conclusion is that by this contradiction θ is not an effectively calculable function.

However, there is an insurmountable problem in that reasoning: diagonalization is a sintactic mistake, because the values of the arguments cannot be equal, since the first one indicates a list of functions, that is to say, an array, and the second a current natural number, since the functions of the list are from the first-order. If we have the pair <5, *5*>, it is a purely notational equality: it denotes the fifth function of the list, with the number *5* as an argument; In fact the pair is $<\varphi_5, 5>$, or best, $< <i, j>_5, 5>$ (the fifth matrix), and the order that has the function (matrix) in the list is not something inherent to it, but assigned with some criterion unrelated to its functionality, for example, the alphabetical order in the rules that defines them, or the numerical order of its array of values.

But even if the unification of arguments is good, the self-reference raises a greater problem: the n-th function with the argument ***n*** again addresses the lack of argument in the second instance. The nth function is again ð, but what is the argument? If we repeat n, we have that ð (n) = ð (ð(n)), and so:

ð (n) = ð (ð (ð (ð (...)))) (infinite regression)

If we do not repeat n, the function would remain without an established argument:

---

[1] This is a step that used to be called impredicative before the hegemony of formalism, because it defines a function based on another defined on a list of which the first forms part. But for the moment we ignore the objection.

$\partial(n) = \partial(\partial(x))$  (open formula)

We leave it to the discretion of the one who reads the option, but in any case, it is not a value of the argument for which the function is not defined, but the argument itself is missing. But this is not a point against the second-order function, but a wrong formation of the function's argument.

Thus interpreted, although the diagonal formula is correct from the point of view of the formal language at a first level of analysis, it is not so if we examine its argument, and this failure in its formation is not consistently remediable. So it is not a first-order function for which θ is not defined, but the question is with the diagonalization method; in this case it is not possible to construct a well-formed self-referential formula.

### 2.2. Contradiction in terms with recursive functions.

Elliott Mendelson [1979] ("Undecidable Problems", page 265), deduce a contradiction from self-reference. From a second order function θ as in the previous point, it defines another, of a single argument:

$\alpha(z) = \mu y(\ \theta(z, z) = 0 \wedge y = y\ )$

In words, it is the minimum value of ***'y'*** for which the function $\varphi_z(z)$ (corresponding to θ (z, z)) is not defined. (The addition of '***y = y***' is only to show that '***y***' is an argument of the function over which we do not put any conditions.)

Since all operations are recursive, and have been applied to the function θ, which is recursive by hypothesis, the function must be recursive. The values that can be deduced from the definition are:

$$\alpha(z) = \begin{cases} 0 & \text{if } \neg(\exists k)\ \varphi_z(z)=k \quad (\varphi_z \text{ is not defined for } z) \\ \text{--} & \text{if } (\exists k)\ \varphi_z(z)=k \quad (\varphi_z \text{ is defined for } z) \end{cases}$$

Note that α is a converse or opposite function with respect to $\varphi_z(z)$, for all z.

Now, α is partial recursive, so it is in the list which is the first argument of θ. Then there exists some index w such that:

$\alpha = \varphi_w$

But in this case:

$$\varphi_w(w) = \begin{cases} 0 & \text{if } \neg(\exists k)\ \varphi_w(w)=k \quad (\varphi_w \text{ is not defined for } w) \\ \text{--} & \text{if } (\exists k)\ \varphi_w(w)=k \quad (\varphi_w \text{ is defined for } w) \end{cases}$$

So,

$\alpha(w)$ is defined ⇔ $\alpha(w)$ is ***not*** defined.

This contradiction is meant to imply that the function θ could not be recursive either partial or total. But the same objection as in point 2.1 holds: diagonalization is a syntactic mistake in this case, for the same reasons (the arguments of the function are of different type, therefore they can never be equal). To visualize it in the arrangement model, in θ (x, x) the first x is actually a matrix of two columns, a set of ordered pairs, and the second x is a natural number. Even if the index with which the matrix is identified coincides with the natural number, it is not possible to argue that a function so formed is of the first order, let alone that it is part of the list that θ is using.

However, the type of reasoning that it uses deserves to be analyzed carefully. This demonstration does not follow the classical canons of the first order logic for the *reductio ad absurdum*. The contradiction does not result from the assumption of recursivity of the function θ, but it is a self-contradictory construction, which in traditional Aristotelian logic is called *contradictio in terminis*, also called oxymoron in contexts broader than logic (philosophy, rhetoric). The function α, which is used for the absurd, has been constructed as a contradictor of θ; no matter what properties may have θ, α contradicts it. Any function that is used, whether recursive or not, if it has as arguments a list of functions, can be able to generate this type of contradictors. But this function cannot produce any misunderstanding in the calculation of the main function, because it has two equal arguments, which are function indexes. That is, it is not a first-order function, the reason why it does not meet the syntactic requirements of the main function.

### 2.3. Contradiction with language of Set Theory.

Since many negative results are based on constructs and demonstrations that are ultimately tributaries of the Russell paradox, it is necessary to analyze this case, although it does not specifically deal with functions or TM.

In his letter to Frege in June 1902 (see van Heijenort [1967], page 124), Russell thus symbolizes the statement of the paradox (using the formal language of Peano):

w = cls ∩ x ϶ ( x ¬∈ x ) . ⇨ : w ∈ w :=: w ¬∈ w.

... that in an updated language would be:

w = { x : x ¬∈ x } ⇨ [ w ∈ w ⇔ w ¬∈ w ]

Russell noted that Frege's semi-formal theory allowed him to violate two rules of classical logic: A) the prohibition of self-reference, justifying because there are predicates that admit it (the predicate "is english", is itself an English predicate, and the word "polysyllabic" is in turn, polysyllabic). It was not differentiated at that time between first-order predicates (which have as arguments only individuals) and second order (whose arguments can be, in turn, predicates), giving rise to its own paradox, and to that of Gelling-Nelson. B) The prohibition of instantiate a variable in the *definiens*, with the term being defined (*definiendum*); this rule was established by the Aristotelian logics, to avoid circularity, but Frege discarded it because he believed that some contextual definitions in mathematics need that substitution.

The relation of belonging is much more ambiguous than that which descends the verb to be. If the arguments are different, the second must be a set, and the coherent interpretation is that the first term satisfies the law of formation of the second, and another that the first is a member of the extension of the second: ***Let*** w = { x : x ¬∈ x }. In a declarative language, the two interpretations coexist. In an imperative language the interpretation is that each member of the first must be part of the extension of the second. And the difficulty lies in that for the three cases the same symbol is used, without adverb modifiers or adjectives as there are in the natural language. Frege tried to remove from logic the ambiguities and the multiple synonymous of natural language, assimilating it to the language of mathematics, timeless and without modality.

We could say that Russell's paradox is more linguistic than set theoretic.

$\{ x : x \neg\in x \}$

... is just another way of denoting the complement of:

$\{ x : x \in x \}$

Both sets are only partitions of the universal set, "the last totality", so that treating them as sets "per se", as new and separate totalities, is contradictory.

If you give a name to the complement, and for the rules of the manipulation of the symbols of the formal language, it is treated as if it were not the negation of a concept, exotic questions are raised, such as:

"The totality of the non-self-belonging sets, is it a member of the set of non-self-belonging sets?"

If you answer affirmatively, the objection arises: "but the members must to be ***only*** those sets that do not belong to themselves".

If answered negatively, the objection is: "but the members must to be ***all*** sets that do not belong to themselves."

The first objection would be legitimate; the second is not, because if a complement contains itself (which is a complement), by the double negation, it returns to the base set (the complement of the complement). Understanding the relation of belonging from one set to another as the satisfaction of the law of formation of the second term, there is no paradox. If it is understood as forming part of the extension of the second, the very notion of self-belonging would be self-contradictory: an object cannot be at once the part and the whole. The complement as a whole collection is not a member of itself, nor of the original set, without any contradiction, because the totality of a partition of the universal set, is not an element pre-existent; is a new construction. The natural language has its own theory of types.

A simpler example:

"The set of objects that are not books is not a book, therefore it is a member of itself."

The complement of the set of books is not a book, but if it is a member of itself, is it the complement of the complement? Then (double negation) would be a book: absurd. Therefore it is not a member of itself. This raises another paradoxical question: taken as a whole, the set of objects that are not books is not a member of the set of all books, but neither of its complement.

The solution is not so complex: by asking whether the whole of a set may or may not be a member of itself, Russell is imaginatively constructing an element that did not exist in the set of books, or in its complement.

For instance:

U* = { 1, 2, 3, 4, 5, 6 }

P* = { 2, 4, 6 }

I* = { 1, 3, 5 }

I* is the complement of P, but the set that has as a single member the set I*, { {1, 3, 5} } is not a member of neither, nor is it in the universal set.

In the version of the paradox with the barber, and with the deliberately primitive language that he uses to construct it (for to approximate to the language of arithmetic), the barber can only be paralyzed by the contradiction:

“Barber shaves x if and only if x does not shave x”

Then:

“Barber shaves barber if and only if barber does not shave barber”.

But if we use the natural language in a normative mode (understood as a “deferred” form of the imperative mode) and in a temporal way (which is not that less scientific, because physics use a mathematical language, which incorporates time, for centuries, and language in normative mode is incorporated in all procedures, so less from Euclid’s algorithm):

“The barber must shave to x, if and only if x has not shaved to x”.

Then, by replacing x with “the barber,” it results:

“The barber must shave the barber, if and only if the barber has not shaved the barber.”

There is no contradiction, no paradox. The result is a bit trivial: if the barber has not shaved himself, he must shave. The Hume’s razor was counterproductive perhaps, in this case[2].

Russell’s solution was the theory of types, and Zermelo’s was to change the Frege’s Comprehension principle:

$\exists A\ \forall x\ (x \in A \Leftrightarrow \varphi(x))$

... by the one of Separation:

$\forall A\ \exists B\ \forall x\ (x \in B \Leftrightarrow (x \in A \land \varphi(x)))$.

If we interpret the relation of belonging as that the first term satisfies the law of the formation of the set in the second term, and we replace the name of the set in the first term by its Gödel’s number, we see how it could transform the paradoxical Russell’s sentence to an undecidable one.

2 I refer to the famous text in which Hume states that there is no way to deduce a normative proposition from a descriptive one (“A treatise on Human Nature”). But the barber is not receiving moral lessons, but an order. And the orders that will be fulfilled in the future, curiously, take the same grammatical form as the moral norms.

If, on the other hand, we interprete belonging as the computation of the law of the formation of the second term, with the Gödel's number of his own negation as the first term ('w' is the negated set), we see how the Halting Problem was devised.

### 2.2.1. Declarative versus normative/imperative language.

If we interpret the predicate "x ∈ w" as simply describing the members of the set w, and then instantiate x with w, the contradiction already mentioned occurs. But if we interpret it as prescribing an action, "add x to the extension of the set w if and only if x is not a member of w", instantiating x with w produces a regressus ad infinitum; if we are using a mixed language, which has declarative and normative elements, the sentence will never be completed. We will expand on this topic later.

## 2.3. The Turing Problem and the Davis' proof.

Alan M. Turing, in [1936], 8, "Application of the Diagonal Process" raises the problem that would later be known as the unsolvability of the Halting Problem. It does not use the formal language of first-order logic. He raises the possibility of a T.M. that given a standard description (S.D.) of any machine, can record the "u" ("Unsatisfactory") symbol if it is circular, or "s" ("Satisfactory") if it is not (if it is "circle-free").

From the way it is raised, it follows that such a machine would be circular itself, so he concludes that no machine that meets the definition is possible. The reason is that it must examine all the preceding descriptions, and when it arrives at the one which corresponds to its own description, it must begin again with all the preceding descriptions; and so ad infinitum.

The reasoning is correct for that particular machine, but it does not follow that it is valid for any possible machine that tries to verify circularity. That is why the Turing demonstration is not found today in university texts and reference books, but the version popularized by Martin Davis [1958], Davis and Weyuker [1983], and which is now known both in the academic world and outside it, as the "demonstration of the insolubility of the Halting Problem," and is attributed almost unanimously to Alan Turing.

While the Turing's demonstration is correct, but informal and restricted to a very particular type of circularity checking machine, Davis' demonstration is much more formal, though not entirely (because he formalizes first-order logic, whereas second-order logic remains in an informal meta language), and pretends to be universally valid: he tries to transform it into a *reductio ad absurdum* of the very concept of validation of finitude.

From the point of view of the logical form, it is a self-contradictory construction and not a demonstration by the absurd, like other cases that we have already analyzed. But the Davis demonstration brings together several elements that make it deserving of special attention.

First, he defines the problem in terms of a recursive predicate, HALT(x, y), and then goes on to analyze the computability of any program written in a generic imperative language (described in detail in section 2 of the book quoted) that satisfies the requirement that HALT be valid if and only if the program 'y' ends with the input 'x'. Then, he constructs a contradictor program, such that the function that computes is identical to the one we analyzed in 2.1. From that construction he ends up deducing a two-way contradiction, of which he records with a brief "But this is a contradiction", but closes the proof without drawing further conclusions. Exactly as we analyze about Mendelson in 2.1.

He would have to use the contradiction to demonstrate the general non-computability of HALT, for which he would need some second-order logical tools, even in non-formal language. But with great cunning, he does not formally present this demonstration, nor does he use metalanguage to supply it, but rather comes to what Wittgenstein called "prose" as opposed to strict proof: he invokes Church's thesis, to replace the lack of a second-order predicate that deprives HALT of any possibility of being computable, nor does it go to a second-order quantifier with reach over HALT.

But Church's thesis does not entail the HALT's complete non computability. It would only support that HALT is not computable for the Gödel number of the contradictor program, with its own code as input.

It is worth dwelling on the added commentary in support of the thesis of insolubility in Davis, Sigal and Weyuker [1994], ch. 4, Sec. 2, p. 69:

---

> In the light of Church's thesis, Theorem 2.1 tells us that there really is no algorithm for testing a given program and input to determine whether it will ever halt. Anyone who finds it surprising that no algorithm exists for such a "simple" problem should be made to realize that it is easy to construct relatively short programs such that nobody is in a position to tell whether they will ever halt. For example, consider the assertion from number theory that every even number > 4 is the sum of two prime numbers. This assertion, known as ***Goldbach's conjecture,*** is clearly true for small even numbers: 4 = 2 + 2, 6 = 3 + 3, 8 = 3 + 5, etc. It is easy to write a program 𝔓 with the language 𝔏 that will search for a counterexample to Goldbach's conjecture, that is, an even number ***n* >** 4 that is not the sum of two primes. Note that the test that a given even number ***n*** is a counterexample only requires checking the primitive recursive predicate
>
> $$\sim\exists x\ \exists y\ (x \le n)\ \&\ (y \le n)\ \&\ [\ \mathrm{Prime}(x)\ \&\mathrm{Prime}(y)\ \&\ x + y = n\ ]$$
>
> The statement that 𝔓 never halts is equivalent to Goldbach's conjecture. Since the conjecture is still open after 250 years, nobody knows whether this program 𝔓 will eventually halt.

---

This Davis' example far exceeds the scope of the original Halting Problem posed by Turing, and also of any formal system, because it implies a search, in an infinite universe, of the semantics of infinite sentences[3]. It is no longer a question of knowing whether an algorithm is circular or not, or whether, for the particular case of the argument provided, it reaches a solution in a finite number of steps, but rather seeks to investigate an infinite space of possibilities. What is insoluble, in that formulation, is the infinity of cases to be examined, not one in particular as was the original problem. Even syntactically it does not agree with the premises of the original problem: the new problem does not need any argument, it is implicitly clear that it must examine every even natural number greater than 4, until one of them meets the condition; no input is needed. And if none meet the condition, it continues ad infinitum. It is clear that for each number ***n*** it is possible to reach the answer in a finite number of steps. But not for all. It is easy for a formal system to detect that the problem of finding a counterexample to Goldbach's

[3] Evidence that the Halting Problem, as proposed by Davis, is a more epistemological problem than mathematical or computational, is that in the first edition [1983], the example was based on Fermat's Last Theorem. Today, there is a demonstration by Andrew Wiles (in 1993 there was already a version) about its veracity, and we know that a program that seeks a solution in the universe of integers greater than zero and with power greater than two would never stop . But that demonstration of Wiles did not affect the computability of the function that Davis had defined.

conjecture is irresolvable (by means of recursive, finite and extensional ways) from its definition, without having to enter in considerations about computability, since the formula that defines it has an unbounded quantifier, and therefore is a non-recursive predicate:

$\exists n \{ \sim\exists x\ \exists y\ (x \leq 2n)\ \&\ (y \leq 2n)\ \&\ [\ Prime(x)\ \&Prime(y)\ \&\ x + y = 2n\ ]\ \}$

The first formula meets the conditions for the Halting Problem:

*Does the Goldbach's tester machine halt with the input **n**?*

But the second formula exceeds them:

*Does the Goldbach's tester machine halt with any input?*

### 2.3.1. Definability, demonstrability and computability.

We will try to show that there is at least a recursive relation that meets the requirements of finitude verification, and that since it is recursive it is effectively calculable according to Church's thesis, and therefore, computable, according to the Turing's thesis.

Using the relations of substitution and demonstrability constructed by Gödel [1931], p. 186, Relationen 27, 30, 31, and 44, 45, in the Peano-Russell's system PR, (for notation issues we will use Mendelson's version[1979], chap. 3, sec. 4, pg. 156, nº (9 b), Sub(y, u, v) and (13 b) Pf(x, y)), we have that:

z=Sub(y, u, v)

…is the Gödel's number of the expression that results from substituting in the expression with the n.G. 'y', the free variable with n.G. 'v' for the expression with n.G. 'u'; and:

Pf(x, y)

…is the relation between a chain of formulas that constitute a deduction, 'x', and the deduced formula 'y'; then we can construct the predicate

IsDef(#Fml, u) ⇔ (Ep) Pf(p, w))

where

w = Sub(#Fml, u, 13)

#Fml is the Gödel's number of the open formula which expresses in the Peano-Russell's system the function φ, 'u' is the substitution value, and 13 is the Gödel's number of the variable with symbol 'x' which is free in Fml formula. So 'w' would simply be the Gödel's number of the closed formula.

Therefore IsDef is valid if and only if the function expressed by the formula Fml is defined for the argument u. And if a Turing machine M computes the values of the function corresponding to Fml, then:

IsDef(#Fml, u) ⇔ Halt(#M, u)

Now, if we calculate the Gödel number of the negation of IsDef, #Neg, and apply it to the formula itself, we have a version of the well-known undecidable sentence:

***IsDef(#Neg, #Neg) is demonstrable in PR***
***if and only if it is demonstrable***
***~IsDef(#Neg, #Neg).***

And since Halt is the equivalent with Turing machines of the predicate IsDef, it is possible to demonstrate with equal rigor that, in any consistent first-order axiomatic theory, the Halt version is also undecidable. Therefore, it is not possible to draw any conclusion other than that, in a theory, to say Turing-Davis, TD, that formalize the Turing machine functionalism:

***Halt(#NegM, #NegM) is demonstrable in TD***
***if and only if it is demonstrable***
***~Halt(#NegM, #NegM)).***

Davis uses the formal declarative language of first-order logic, but the proper axioms of Turing's machines theory are not made explicit, and he attributes the semantic values of ***true*** or ***false*** to the Halt predicate as if he could jump from a formalism without axioms to some model, without also defining the interpretation in Theory of Models. But let us suppose for a moment that the result that he attempts to rescue from his demonstration is valid, that is, that Halt is not computable. Since IsDef is its equivalent in the PR language, this would mean that IsDef is not effectively calculable, and therefore, that it is not recursive. Since it is constructed on the basis of the substitution function and the provability relation, which are primitive recursive, this is not possible. Therefore Halt is computable to the same extent and in the same domain where IsDef is effectively calculable.

## § 3. Inverse Halting Problem.

Behind the Halting Problem are several incompatible problems.

We can identify three types of possible infinite computations, according to their cause:

A) By circularity.

B) Some variable(s) have infinite range.

The last case can be divided into two:

B.1. The domain is finite in the definition, but there are variables that in the invariants have no upper or lower bound.

B.2. The definition formula of the problem in first order language contemplates an infinite domain.

They are incompatible because the solution that could have one does not imply the solution of the others, and even any solution proposed to solve them together, can solve one and mask and/or aggravate the others.

Problem A) is detectable by simple parsing if working in the realm of recursive functions. If it is Turing machines, or programs written in imperative languages analogous to that described by Davis, it would not be too difficult to prove that if an invariant repeats the values of its variables, its trace, or snapshot (as Davis and Weyuker [1983] ], Chapter 2, Sec. 4) is periodic.

The problem B.1) by calculating the invariants of each sentence.

Problem B.2) is unsolvable, but it has nothing to do with computability: it is a matter of first order logic, and it is outside the scope of recursive definitions. A problem that in its approach uses formulas that express non-recursive relations, obviously is not solved in general with recursive methods. It is not necessary to prove incomputability, since the definition itself is not recursive.

There is a fourth problem, which has nothing to do with infinity, but with the type of language used to reason about the problem. As we saw in 2.2.1, a contradiction in declarative language tells us nothing about what would happen with an implementation in an imperative language. In a metalanguage we could see that, since the contradiction is due to circularity in the definition, in an imperative language would become an infinite regress.

If we consider the inverse Halting Problem, that is, in what ranges an imperative implementation becomes infinite, the problem is reduced to evaluating a system of two equations: the output condition of a cycle, and the previous invariant (the precondition). If the algebraic problem has no solutions, the computation is infinite. That is, it may have a precise answer.

## § 4. Conclusions.

1) In the domain of recursive functions the circularity is detectable as a syntactic failure in the construction of the formulas, so that the problem of the infinite loops of the Turing machines in the recursive functions is not irresolvable. The ties of Turing algorithms and similar languages could very well be avoided if each of them were constructed through transcription into their imperative language from a general recursive function expressed in a formal first-order language. And it can be done automatically.

2) If each computer procedure corresponds to the transcription in imperative language from a primitive recursive function, expressed in the language of first order logic plus the “axiomatic” definitions of Gödel’s framework, something similar to the *desideratum* proposed by Sir Tony Hoare [2003]: A Verifying Compiler, could be reached. The reason is that these functions have a critical casuistry to verify the calculation. In the same way that the canonical bases of a linear vector space allow us to verify if a linear transformation leads from one space to another, by simply verifying that the canonical basis of the domain leads to the canonical basis of the range.

3) In the uncertainty situation of type 4, of the first part (on a function defined with the use of the minimization operator, whose equation we don’t know if it has solutions), if the minimization operator is not bounded, it is not a recursive definition. It is an epistemological question, which concerns to the historical development of the knowledge of mathematics in a given context, and not to the mathematical entities themselves. The same applies to programs that calculate these functions.

4) If, as Huizing, Kuiper, and Verhoeff [2010] argue, “Halting Still Standing”, it can be attributed to the linguistic presuppositions underlying natural language rather than to the logic or computability of functions. The construction of a self-contradictory sentence in a formalized or semi-formalized language can always give rise to widespread skeptical conclusions to the whole field of the discipline that uses that language. In this sense, Halting closely resembles the Berry-Richard paradox: “*let N be the first non-definable ordinal; if it is proved that it is effectively not definable in a language, the very statement of the problem is transformed into a definite description of N (the first non-definable ordinal); so, it is definable after all*”. Chaitin [1990] formalizes the idea. In other words, if the blind spot of a Halting function is computable in some language, it is always possible to raise it by constructing another blind spot, by means of some different Gödelization, for example (see Hehner [2015], paragraph “How to compute unlimited halting”). But that does not make the whole function incomputable.

It is possible to make a final distinction between semantics in the operative sense, as an extension of concepts, predicates and functions, on the one hand, and intentional (mental, psychic) semantics, on the other, the pure meaning of propositions, which they can be lightened but not fixed.

Doubts and concerns that may affect the academic field on the basis of the limitations of intentional semantics should not affect the ability to construct fully operative and effective formalisms. Contradictory constructions, along with the idea of common sense that a universal verification procedure is impossible, is the basis of skepticism. But it is not a matter of constructing a universal verifier of all possible propositions, but of a second-order function which validates that first-order functions are defined in the mathematical sense. And in turn, a translator of functional formulas to programs in imperative language. And this is perfectly feasible, we hope.

Acknowledgement: This work is based mainly on the thesis that professor Eric Hehner, from the University of Toronto, has been raising for more than thirty years. I have also had the privilege of receive his encouragement, corrections and orientation.

Buenos Aires, February 2017.

---